# Dissipation of the excess energy of the adsorbate- thermalization via electron transfer

Paweł Strąk, Konrad Sakowski and Stanisław Krukowski

Institute of High Pressure Physics, Polish Academy of Sciences, Sokołowska 29/37, 01-142 Warsaw, Poland

**Abstract**

A new scenario of thermalization process of the adsorbate attached at solid surfaces is proposed. The scenario is based on existence of electric dipole layer in which the electron wavefunctions extend over the positive ions. Thus the strong local electric field exists which drags electron into the solids and repels the positive ions. The electrons are tunneling conveying the energy into the solid interior. The positive ions are retarded in the field, which allows them to loose excess kinetic energy and to be located smoothly into the adsorption sites. In this way the excess energy is not dissipated locally avoiding melting or creation of defects, in accordance with the experiments. The scenario is supported by the *ab intio* calculation results including density function theory of the slabs representing AlN surface and the Schrodinger equation for time evolution of hydrogen-like atom at the solid surface.

## I. Introduction

Energy dissipation during surface reaction is an important part of the studies of surface reaction and catalysis. Due to its importance to industrial technologies where heterogeneous catalysis is the core of the estimated 80% processes[1], the investigations of heterogeneous catalysis attract constant attention of the researchers. A glaring example of the importance of the subject is the discovery of Haber-Bosch synthesis of ammonia, a hidden pillar of the modern civilization, enabling production of the vast quantity of food and avoiding famine in the global scale. Unraveling of the mechanism of this reaction was awarded by the 2007 Nobel Prize for G. Ertl [2].

The complexity and sheer scale of difficulties of the subject is comparable to its importance. The number of issues in surface reactions is staggering, therefore the subject of the present study will be limited to excess energy dissipation and thermalization of the surface reaction products at the surface, adatoms, admolecules or radicals[3,4]. The energetic effect of adsorption at surfaces frequently reaches several electronvolts[3-10]. The examples include dissociative adsorption of oxygen at Pd(100) surface leading to energy gain of 2.6 eV[3], nitrogen at AlN(0001) surface - 6.0 eV[5], oxygen at Ag(001) surface – 1.65 eV[6]. Similar effects arises due to adsorption of hydrogen at Pd(100) and Pd(111) surfaces[7,8], fluorine at Si(111) surface[8], chlorine at $TiO_2$ surface[9] or hydrogen at GaN(0001) surface[10]. The large adsorption energies are therefore observed for processes involving a number of molecules, including $O_2$, $H_2$, $N_2$, $Cl_2$ or $F_2$ at surfaces of various types, including metals, palladium, aluminum or silver, semiconductors such as silicon, gallium nitride or aluminum nitride or even oxides such as titanium dioxide. These energies are attained in excess to relatively large dissociation energies of the molecules such as 9.8 eV for molecular nitrogen, 5.2 eV for molecular oxygen or 4.46 eV for molecular hydrogen[11]. These gains are due to strong bonding between chemically active surfaces where dangling bond states saturation generates large chemical energy gains. The excess energy has to be dissipated to realize localization of the atoms in the surface sites to realize chemical bonding.

The excess energy dissipation model during surface collision should describe process that is fast, efficient, and nondestructive to the surface, despite its magnitude. The two basic scenarios were identified: dissociative adsorption in which the products are attached at the surface sites, or the abstractive process in which only one atom is attached at the surface whereas the second is ejected in the vacuum carrying out the energy excess. The latter process is observed during adsorption of $F_2$, $Cl_2$ and $Br_2$ at silicon surfaces[8].

The dissociative process entails kinetic energy dissipation as the energy is kept within the surface-solid system. In the effort to elucidate the process several possible energy dissipation channels was proposed[12, 4], including creation of electron-hole pairs[13-15], excitation of phonon bunches[3,16], or persistence of "hot" adatoms, i.e. high energy atomic products at the surface[6,7,16-19].

The experimental evidence does not confirm any of these hypotheses, giving mixed responses, depending on the adsorbate and the surface. The semiclassical phonon excitations model explains successfully results of nonreactive noble gas atoms with surfaces[20]. The reactive species adsorption cannot be understood within such models[21]. The e-h model describes some aspects of H(D) adatom energy dissipation, mostly for light adatms such as hydrogen[14, 15]. It was been found however that for scattering of heavier atoms and molecules, such as molecular nitrogen at W(110) surface, the e-h contribution plays minor role[12].

"Hot" adatoms were clearly identified for number of surface reactions, such as oxidation of Al(111) surface[17,19-21] or adsorption of oxygen at Pt(111) surface[18]. The experimental evidence from scanning electron microscopy (STM) for a number of surface reactions[6]. *Ab intio* molecular dynamic studies (ABMD) confirmed existence of such energy dissipation channel[7]. It was also found that the hydrogen adatom could travel about 7 Å at the Pd(100) surface and not more for heavier species such as oxygen[7]. The STM investigations provide different picture[6]. For nonionic crystals surfaces such as Pt(111)[18], Cu(110)[22] these distances are small, of single lattice constants, i.e. below 1 nm. On the other hand, the distances covered by "hot" adatoms at the surfaces of nonconductive crystals, such as $TiO_2$ (110) was much higher, about 2.6 nm[9]. Notably, the distance measured for adsorption of oxygen at Al(111) surface was even higher, close to 8 nm[17,20]. It is still not clear whether such difference could not be attributed to oxidation of Al(111) surface, thus confirming such large distance related to electric conduction of the crystal. It has to be understood that the "hot" atom scenario does not provide any answer how the excess kinetic energy is dissipated, the only result is that the adatoms are separated due to parallel motion. Nevertheless the energy is dissipated, thus the answer has to be sought further[13].

Thus the two existing choices are phonon and electron related dissipation. The phonon scenario entails scattering with the atoms at the impingement point and along the adatom path. This may include such approaches as local density friction coefficient approximation (LDFA)[23,24] or the generalized Langevin oscillator model (GLO)[25]. Nevertheless, all these approaches entail kinetic energy transfer to the atomic nuclei, by direct interaction only. As the number of these atoms, in the case the shortest path, measured for the electric conductors, is below 10, the kinetic energy possibly transferred to any of them is of order of 1 eV, that would entail melting of the segment of the surface. This, in case of the nitrides, would entail creation of nitrogen molecules and their escape from the surface. Thus, the intensive decomposition of AlN or GaN due to their exposure to molecular nitrogen should be observed. No such effect is observed.

It has to be noted that the impingement of the molecules at the surface has peculiar angular properties. The attraction by the surface is directed perpendicular to the surface, thus most of kinetic energy is associated with the momentum component perpendicular to the surface, which is several electronvolts. Thus the kinetic energy fraction associated with the momentum parallel to the surface is of order of the thermal motion energy, i.e. of order of 0.3 eV. The latter may contribute to "hot" atoms motion at the surface. The first should be transferred to one or at most three atoms, the closest to impingement point. Then this energy would be transferred to the neighbors by phonons. Again the local melting would be inevitable, the phenomenon not observed. Worse, impingement of single nitrogen atoms, in plasma MBE processes, especially with use of bright plasma would entail melting and nitrogen escape[26]. No such process is observed in MBE growth of gallium, aluminum or indium nitride layers[27]. Additionally, the phonon models do not explain close connection between electrical conductance and the magnitude of the adatom shift at the surface. Therefore phonon models cannot explain the observed dissipation of kinetic energy in the reactive species adsorption.

The second possible choice is the energy transfer to electronic degrees of freedom. The typical scenario assumes creation of the e-h pair within simplified rate model, with no detailed explanation of the mechanism of the event[14]. More detailed studies invokes emission of electrons observed during adsorption of NO molecules at low work function metal surfaces[28]. The latter observation proves that the electrons are involved in the energy dissipation, but it does not indicate that e-h pairs are created. Moreover, recent assessment of the phonon and e-h energy dissipation processes in case of adsorption of $N_2$ at W(110) and N at Ag(111) surfaces shows that e-h pair creation represents 10% of the energy dissipation so that the phonon emission is more effective channel[12]. As the result was identical for both molecules and the atoms, this excludes e-h pair creation as the viable channel of energy dissipation at metal surfaces.

It is therefore necessary to look for other electron associated channel of effective kinetic energy dissipation that is not related to e-h pair creation. This scenario is dated back to jelly model by Lang and Kohn[29]. They identified dipole layer at surface of metal and associated the existence of negative charge outside with the work function. The recent investigations of the work function of nitrides provide new insight on the band structure and electric potential. The model described below present new scenario based on electron transfer from the adsorbate to the solid. The model could describe the thermalization process without possibility of local melting, bringing the excess energy into the deep interior of the solid and retardation of the charged impinging species by electrostatic potential barrier at the surface.

## II. Calculation methods

*Ab initio* calculations used in the determination of the data presented in this paper were obtained using two approaches. First part was devoted to simulation of AlN(0001) surface used SIESTA[30-32]) shareware, with numeric atomic orbitals, having by finite size predetermined support. The ATOM program employing all-electron calculations, from code authors was used to generate pseudopotentials for Al and N atoms.[33] In k-space integration a Monkhorst-Pack grid (5x5x1) [34] with the Perdew, Burke and Ernzerhof for solids (PBEsol) devised exchange-correlation functional in Generalized Gradient Approximation (GGA) was used.[35,36] The plane wave cutoff for the properties calculations was 275 Ry. SCF loop was terminated when the maximum difference between the output and the input of each element of the density matrix was below $10^{-4}$. The positions of atoms were modified for the forces on every single atom above 0.005 eV/Å. The AlN *ab initio* lattice constants were a = 3.116 Å, c = 4.974 Å remaining in acceptable agreement with the experimental data: a = 3.111 Å, c = 4.981 Å.[37] Ferreira et al. LDA-1/2 and GGA-1/2 correction schemes, giving proper band gap energies, effective masses, and band structures provided bandgap energy $E_g$(AlN) = 6.16 eV. [38,39] in a reasonably good agreement with the following low-temperature experimental data $E_g$ (AlN) = 6.09 eV.[36] The electronic properties were obtained in modified GGA-1/2 scheme for which positions of atoms and a periodic cell were first obtained using PBEsol exchange-correlation functional.

The second part of the calculation was to solve time dependent quantum mechanical evolution of the atom in the potential profiles derived from *SIESTA* data. The solution was obtained using finite element solver for Schrodinger equation developed by the authors. Thus the single electron problem was solved. The solution was obtained using nonuniform grid, dense in the vicinity of the high field region. The probability of tunneling was obtained by integration of the probability of finding the electron behind the potential energy maximum, assuming that initially the electron wavefunction is limited to the part before the barrier.

## III. Results

As an example for simulations of the thermalization during adsorption, the polar Al surface of aluminum nitride, i.e. AlN(0001) surface was used. The surface was selected because AlN is relatively hard material, of wide direct bandgap, thus the competitive thermally activated processes have smaller importance. Additionally aluminum nitride is grown from the vapor phase by direct sublimation via transport of atomic Al and molecular $N_2$ vapors.[41] The surface reaction entails adsorption of these species, leading to their attachment in atomic or molecular form.[42] The reaction between liquid Al and $N_2$ is highly exothermic, leading to rapid dissociation at the metal surface.[43] Recent ab intio simulations of $N_2$ adsorption at AlN(0001) surface indicated that at low N coverage, the reaction is dissociative with the energy gain of 6.05 eV/molecule.[44] Thus the excess energy is considerable, and despite that no meltback is observed and the molecular nitrogen escape is observed. Thus the thermalization processes are extremely effective, that excludes the above discussed "hot" atoms and e-h creation models.

Therefore as the best example of the *ab initio* calculations, the band structure of aluminum nitride surface, the electric potential profile across AlN slab, obtained from SIESTA code, is presented in Figure 1.

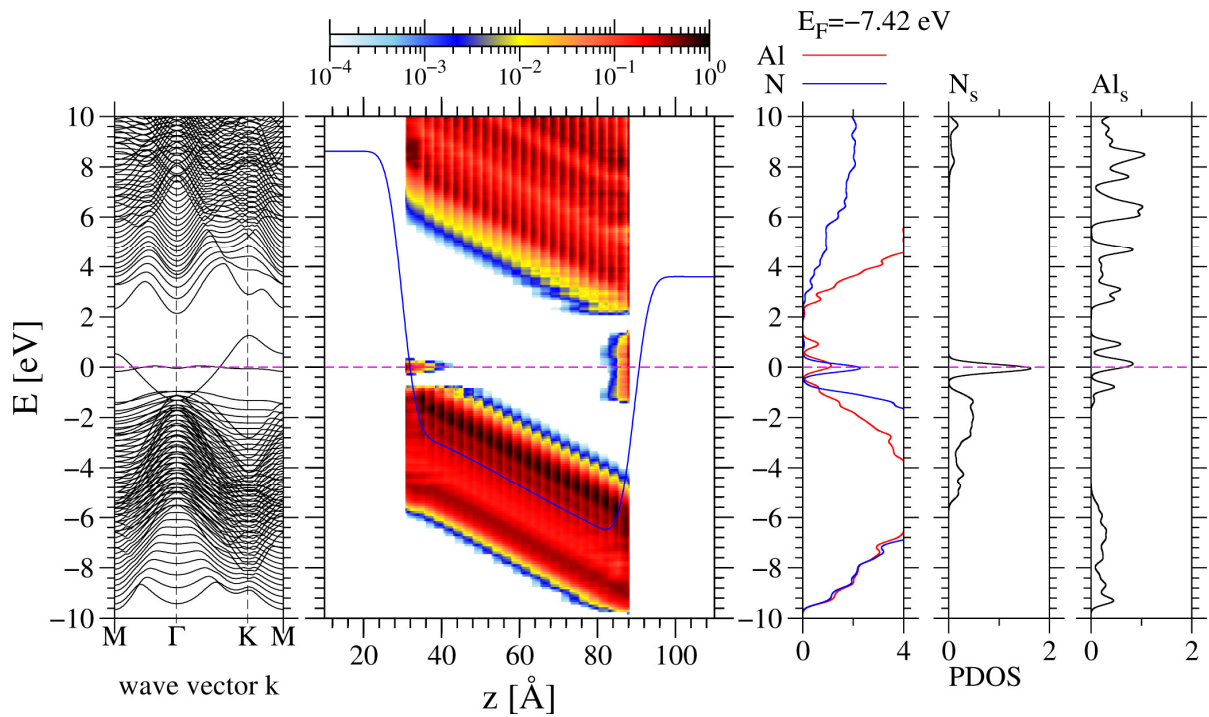

Figure 1. Band diagram in momentum (left) and position space (middle) of the WZ 24 double atomic layer (DALs) of semi-insulating AlN slab and the density of states (DOS) of surface Al atoms (right). The line superimposed on the position space diagram is electric potential, averaged in the plane parallel to surface of the slab, plotted in units of electron energy.

As it is shown at both polar AlN surfaces the electric potential suffers the drop of about 10 eV over the distance of 2 nanometers. Thus the strong electric field arises, which is identified as caused by the electron wave functions extending outside the atomic nuclei[29]. Thus, the impinging species encounters region of electric field which drags their electrons into the solid interior. In fact, the force originates from repulsion of the electrons outside and attraction by the positive ions inside. In order to investigate the process, the model potential was created for the case of hydrogen atom that accounts the slab field and the attraction by the hydrogen nucleus. The potential is presented in Figure 2.

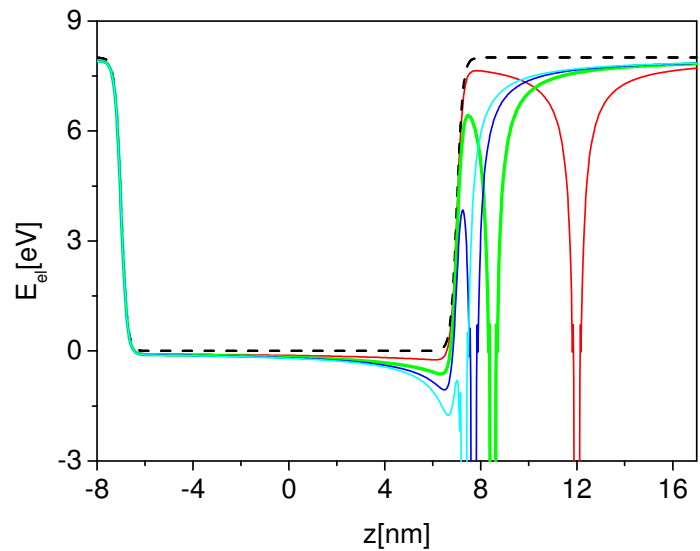

Figure 2. The model potential of the hydrogen atom and the finite slab of the solid, for selected several distances between the atom and the surface. The potential is expressed as electron energy assuming the following distances between the atom and the surface: red line - 5 nm, green line - 1.5 nm, blue line - 0.7 nm, cyan line - 0.3 nm. The dashed black line represents potential of the slab. The solid green thick line represents the distance 1.5 nm for which time dependent Schrodinger equation calculations were made.

The possibility of electron transfer was investigated by solution of time dependent Schrodinger equation. The obtained time dependent wavefunction is presented in Figure 3 where the temporal evolution of the spatial distribution of electron density is drawn for several selected times. The selected adatom-surface distance for the time dependent simulation is 1.5 nm. The initial state of the electron is represented by the s wavefunction which in the logarithmic coordinates is represented by straight line. Naturally the results contains some noise, nevertheless the results clearly demonstrates the tunneling across the barrier which is visible on the right part of diagram in Figure 3. The initial state was selected so that the energy in the hydrogen potential is close to the bottom level energy of the solid slab.

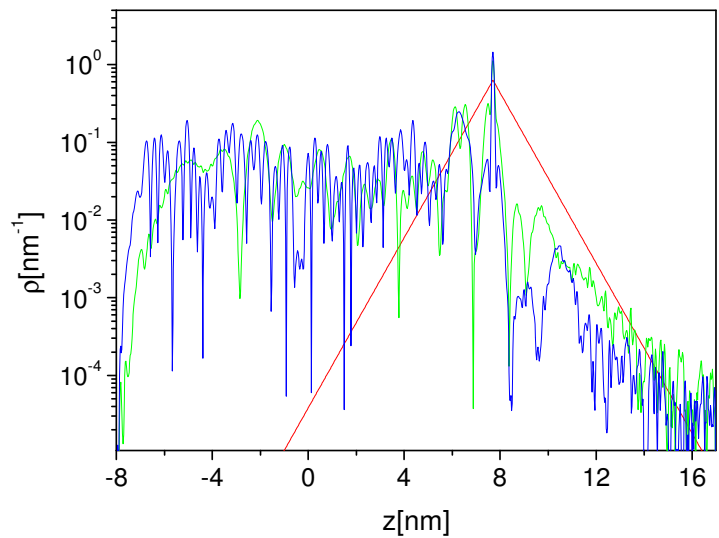

Figure 3. Electron density distribution snapshots, obtained for the atom-surface distance d = 1.5 nm: t = 0 - red line; t = $10^{-15}$ s - green line; t = 2 *$10^{-15}$ s - blue line. The initial state at t = 0 is s-state.

As it was shown by the electron density change in time, the fraction of the probability distribution of the electron behind the barrier becomes considerable in time much shorter than t ~ $10^{-14}$ s. That is presented more effectively by the temporal evolution of the probability of transition shown in Figure 4. The probability was obtained by summation of the density of the wavefunction in the region behind the potential maximum, which for the simulated case corresponds to z < 7.41 nm.

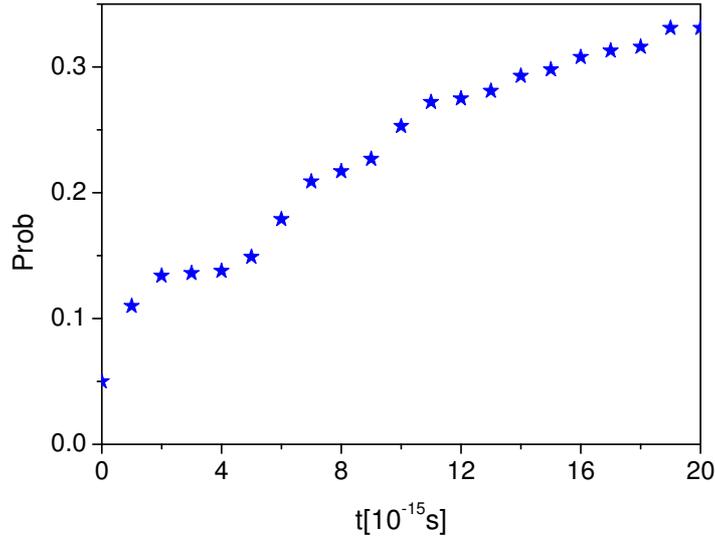

Figure 4. Probability of tunnelling from the hydrogen-like atoms to the solid slab in function of time for the atom-surface distance d = 1.5 nm.

From these simulations it follows that considerable probability was attained after $\tau_{jump} \approx 2 \cdot 10^{-14} s$. For longer time the transition saturates and only the fluctuations are observed. This result has to be compared with the duration of the scattering event of the surface with the atom having typical thermal velocity.

An assessment of the duration of collision process may be made using normal component of the velocity of nitrogen atom at T = 1300K. At this temperature the thermal velocity normal component may be obtained from equipartition principle as $v = \sqrt{\frac{k_B T}{2 M_N}} \approx 600 \, m/s$. Assuming the collision process interaction range $d = 5 \text{Å}$, we obtain the collision time $\tau_{coll} \approx 8 \cdot 10^{-13} s$. The tunneling time is about two orders of magnitude shorter than duration of the typical atom-surface collision process and could be treated as instantaneous. This time scenario confirms the prediction of the possible role of electron transfer and the retardation of the positively charged adatom/admolecule at the final stage of collision process. Thus the electron transfer is possible channel of kinetic energy dissipation at the surfaces.

The mechanism of retardation may be presented by the drawing of the energy in function of the distance from the surface of the solid.

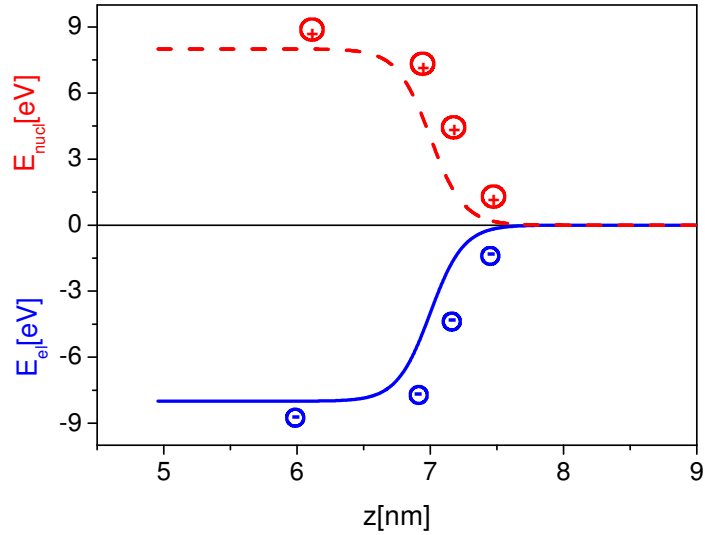

Figure 5. The potential energy of the electron (blue solid line) and the positive ion (red dashed line) derived from AlN slab results of *ab initio* calculations presented in Figure 1.

As it is shown the electron could be accelerated by the electric field. At the same time the positively charged ion is retarded. Depending on the ionization moment, the electron energy gain and equivalently, the ion energy loss may reach 8 eV. Thus the described here new scenario of the thermalization assumes that the excess kinetic energy is conveyed by the electron deep into the solid interior, while the positively charged ion is retarded at the surface so that smoothly arrives at the adsorption site. The electron is shifted to conduction band where it looses the excess energy by phonon emission i.e. standard intraband relaxation process. The difference is that the latter process is not local, but includes large number of atoms in the lattice, thus avoiding drastic local temperature increase and the possible melting or generation of the defects.

**IV. Summary**

In summary, the new scenario of thermalization of kinetic energy of the adsorbate was proposed, based on the existence of dipole layer at the surfaces of the solids. Due to strong local electric field related to the existing dipole layer, the arriving species are ionized by the field emission via tunneling of the electron into the solid interior. In the tunneling event the electron acquires energy of a few electronvolts conveying the energy excess into the solid. Subsequently the electron is thermalized inside the solid interior travelling long distance and emitting the phonons along its path. Thus the energy dissipation occurs over the large distance avoiding local generation of large quantities of heat and creation of lattice defects.

The dipole related surface electric field plays also the additional role in thermalization. As the atomic/molecular ionized species are positively charged, the interaction with the dipole related potential cause drastic retardation of the positive ions towards the surface thus loosing the excess energy. In the result the ions slowed down are attached smoothly at the surface of the solid avoiding effect related to large excess kinetic energy. Thus the thermalization process is effective, not leading to creation of defects, as observed in experiments.

**Acknowledgement.**

The research was funded by Polish National Science Center by grant: DEC-2015/19/B/ST5/02136. The calculations reported in this paper were performed using computing facilities of the Interdisciplinary Centre for Mathematical and Computational Modelling of University of Warsaw (ICM UW).